# Surpassing the classical limit of microwave photonic frequency fading effect by quantum microwave photonics

## Authors

Yaqing Jin[1,3], Ye Yang[2,6], Huibo Hong[1,3], Xiao Xiang[1,3], Runai Quan[1,3], Tao Liu[1,3], Ninghua Zhu[2,4,5], Ming Li[2,4,5#], Ruifang Dong[1,3*] & Shougang Zhang[1,3*]

## Affiliations

*[1] Key Laboratory of Time and Frequency Primary Standards, National Time Service Center, Chinese Academy of Sciences, Xi'an 710600, China*
*[2] State Key Laboratory on Integrated Optoelectronics, Institute of Semiconductors, Chinese Academy of Sciences, Beijing, 100083, China*
*[3] School of Astronomy and Space Science, University of Chinese Academy of Sciences, Beijing 100049, China*
*[4] School of Electronic, Electrical and Communication Engineering, University of Chinese Academy of Sciences, Beijing 100049, China*
*[5] Center of Materials Science and Optoelectronics Engineering, University of Chinese Academy of Sciences, Beijing 100190, China*
*[6] The 29th Research Institute of China Electronics Technology Group Corporation, Chengdu 610029, China*
*Corresponding author [#]ml@semi.ac.cn, *dongruifang@ntsc.ac.cn *szhang@ntsc.ac.cn*

**Abstract:** ***With energy-time entangled biphoton sources as the optical carrier and time-correlated single-photon detection for high-speed radio frequency (RF) signal recovery, the method of quantum microwave photonics (QMWP) has presented the unprecedented potential of nonlocal RF signal encoding and efficient RF signal distilling from the dispersion interference associated with ultrashort pulse carriers. In this letter, its capability in microwave signal processing and prospective superiority is further demonstrated. Both the QMWP RF phase shifting and transversal filtering functionality, which are the fundamental building blocks of microwave signal processing, are realized. Besides the perfect immunity to the dispersion-induced frequency fading effect associated with the broadband carrier in classical microwave photonics, a native two-dimensional parallel microwave signal processor is provided. These demonstrations fully prove the superiority of QMWP over classical MWP and open the door to new application fields of MWP involving encrypted processing.***

Microwave photonics (MWP), which deals with the generation, processing, control, and distribution of high-speed radio frequency (RF) signals using photonic techniques[1], has attracted increasing interest as it can achieve functions that are very complex or even not possible using purely electronic technology[2-7]. Its application has been widespread in numerous areas such as broadband wireless communication networks, radar[8], instrumentation, and sensor networks[9]. A microwave photonic system should typically include a laser source as the optical carrier, an electro-optic modulator (EOM), an optical signal processor, and a photodetector (PD). The microwave or RF signal is first modulated onto the optical carrier by the EOM, and the modulated optical carrier is then processed by the optical signal processor. With the PD, the processed optical signal is detected for the RF signal recovery. However, for the scenarios with ultra-low power and low signal-to-noise ratio, the current microwave photonic systems cannot be applied. Furthermore, in various MWP applications using broadband sources as the optical carrier, the processed RF signal also suffers severe dispersion-induced frequency fading, which becomes a more and more important problem. Besides, although the processing bandwidth of the optical technology can be as large as hundreds of THz, the bandwidth of the classical MWP system is ultimately limited by the utilized microwave devices for further processing[10,11].

To break the above bottlenecks, the scheme of single-photon microwave photonics (SP-MWP) was recently proposed[12]. Based on a superconducting nanowire single photon detector (SNSPD) and a successive time-correlated single photon counting (TCSPC) module, the SP-MWP signal processing system with phase shifting and frequency transversal filtering was demonstrated. An ultrahigh optical sensitivity down to −100 dBm has been achieved and the signal processing bandwidth is only limited by the timing jitter of single-photon detectors, which can be as low as sub-3 picosecond[13]. Nonetheless, the dispersion-induced frequency fading associated with the carrier bandwidth still remains.

With the rapid growth of photonic quantum technology, it has promised enhancements to a vast range of fields including navigation and timing[14-17], secure communications[18-20], imaging and sensing[21,22], and quantum computing[23]. Recently, quantum microwave photonics in radio-over-fiber (QMWP-RoF) systems has been demonstrated using an energy-time entangled biphoton source as the ultrashort optical pulse carrier combined with the SPD technique[24]. Due to the strong quantum temporal correlation between the energy-time entangled photon pairs[25,26], the QMWP provides the unprecedented capability of nonlocal RF modulation, which also realizes improved spurious-free dynamic range (SFDR) and strong resistance to dispersion interference in RoF systems. Together with the advantages of ultra-weak detection and high-speed processing rendered by the SPD technique, the QMWP method offers a bright potential for exploring new possibilities in modern communication and networks.

In this paper, we present the first QMWP signal processing system to demonstrate its superior capability over classical MWP. Both functions of RF phase shifting and multi-tap transversal frequency filtering have been realized, which are the fundamental building blocks of microwave signal processing. In contrast to the classical MWP, the significant dispersion-induced frequency fading associated with the carrier bandwidth can be perfectly eliminated, which thus overcomes the bandwidth limit in classical broadband MWP[1]. Besides, as both of the entangled signal and idler photons can be employed as the optical carrier, a two-dimensional parallel microwave signal processer is naturally provided with each photon carrier taking independent phase shifting and transversal filtering characteristics. These demonstrations of QMWP signal processing fully prove the superiority of QMWP over classical MWP and open the door to a new application field of microwave photonics such as encrypted processing.

## Results

### Principle

To illustrate the core principle involved in the QMWP signal processing, the schematic diagram for realizing the nonlocal RF signal mapping, phase shifting and 3-tap transversal filtering is shown in Fig. 1.

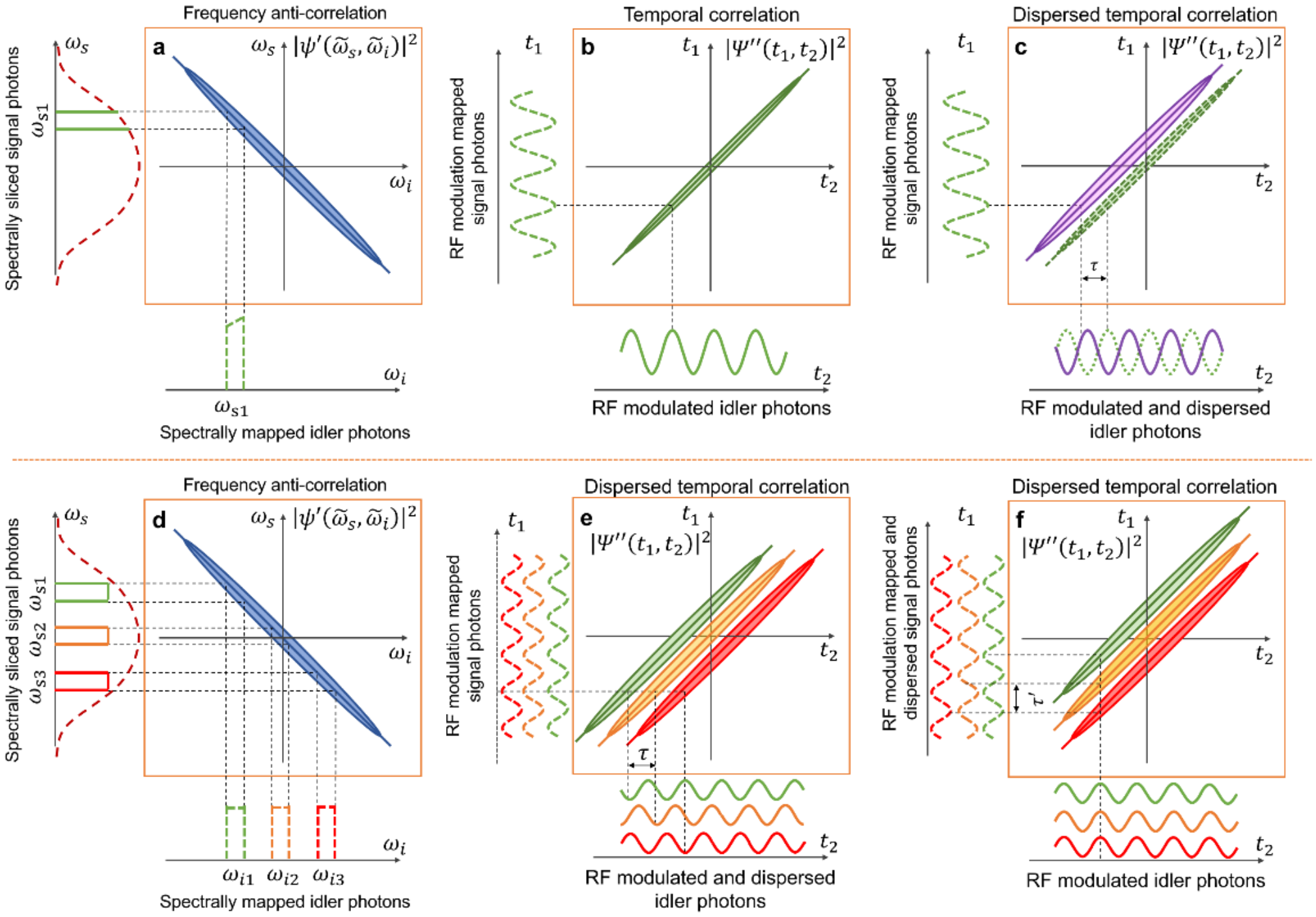

**Fig. 1 Principle diagrams of the QMWP-based nonlocal RF phase shifter and transversal filter. a** Illustration of the nonlocal spectral mapping based on the joint spectrum of frequency anti-correlated photon pairs. Because of the strong frequency anti-correlation, the spectral slicing on the signal photons will be nonlocally mapped onto the idler photons. By adjusting the passband channel of the signal photons, the corresponding passband channel of the idler photons is changed in the opposite direction. **b** The joint temporal distribution profile of the spectrally sliced photon pairs. Due to the strong temporal correlation, the intensity modulation on the idler photons will be nonlocally transferred to the signal photons. **c** The visualized evolution of the temporal correlation distribution and modulation when a dispersive element is applied to the idler photon, which is dispersed both in width and offset. When the modulation is applied, the phase shift due to dispersion-induced offset will results. **d** The nonlocal mapping on the idler photons when the optical slicer with multiple passbands is applied. **e** The visualized evolution of the temporal correlation distribution and modulation with the multi-passband slicer on the signal photons while the dispersion on the idler photons. **f** The visualized evolution of the temporal correlation distribution and modulation with the dispersion applied to the signal photons.

The energy-time entanglement refers to the signal and idler photons having the quantum properties of frequency anticorrelation and temporal correlation. When the CW pumped spontaneous parametric down conversion (SPDC) is utilized to generate the two-photon state, its spectral wave function can be given by

$$\psi(\widetilde{\omega}_s, \widetilde{\omega}_i) \propto \delta(\widetilde{\omega}_s + \widetilde{\omega}_i)\mathrm{sinc}\left[(\gamma_s\widetilde{\omega}_s + \gamma_i\widetilde{\omega}_i)\frac{L}{2}\right], \quad (1)$$

where $\widetilde{\omega}_{s(i)} = \omega_{s/i} - \omega_{s/i,0}$ is the angular frequency detuning of the signal (idler) photon from its center angular frequency $\omega_{s(i),0}$. $\gamma_{s(i)}$ is the phase matching parameter; and $L$ is the length of the SPDC crystal[27]. From Eq. (1), it is clearly seen that the signal and idler photons are frequency anticorrelated. The schematic diagram of the joint spectrum of frequency anti-correlated photon pairs is shown as Fig. 1a. The sinc function is the crystal's phase-matching function, whose bandwidth denotes the single photon spectral bandwidth. When an optical slicer at $\widetilde{\omega}_F$ with a spectral bandwidth of $\sigma_F$ is applied to the signal photons, which can be described by $H(\widetilde{\omega}_s) \sim e^{-(\widetilde{\omega}_s - \widetilde{\omega}_F)^2/(2\sigma_F^2)}$, the two-photon spectral wave function is transferred to

$$\psi'(\widetilde{\omega}_s,\widetilde{\omega}_i) \propto \delta(\widetilde{\omega}_s+\widetilde{\omega}_i)\mathrm{sinc}\left[(\gamma_s\widetilde{\omega}_s+\gamma_i\widetilde{\omega}_i)\frac{L}{2}\right]e^{-\frac{(\widetilde{\omega}_s-\widetilde{\omega}_F)^2}{2\sigma_F^2}}. \quad (2)$$

When $\sigma_F$ is far narrower than the crystal's phase-matching bandwidth, the two-photon spectral wave function can be reduced to

$$\psi'(\widetilde{\omega}_s,\widetilde{\omega}_i) \propto \delta(\widetilde{\omega}_s+\widetilde{\omega}_i)e^{-\frac{(\widetilde{\omega}_s-\widetilde{\omega}_F)^2}{2\sigma_F^2}}, \quad (3)$$

Under such condition, the spectral wave function of the idler photons can be deduced as

$$\varrho_c(\widetilde{\omega}_i) = \int d\widetilde{\omega}_s\, \psi'(\widetilde{\omega}_s,\widetilde{\omega}_i) \propto e^{-\frac{(\widetilde{\omega}_i+\widetilde{\omega}_F)^2}{2\sigma_F^2}}. \quad (4)$$

Therefore, as depicted in Fig. 1a, the spectral slicing of the signal photons at $\widetilde{\omega}_F$ is nonlocally mapped onto the idler photons at $-\widetilde{\omega}_F$.

The two-photon temporal wave function can be deduced from the Fourier transformation of Eq. (3):

$$\Psi'(t_1,t_2) \propto e^{-\frac{\sigma_F^2}{2}(t_1-t_2)^2-i\widetilde{\omega}_F(t_1-t_2)}, \quad (5)$$

where $t_1$ and $t_2$ represent the temporal coordinates of the signal and idler photons respectively. It is clearly seen that the spectrally sliced photon pairs are temporal correlated and their joint temporal distribution is sketched in Fig. 1b. When the RF signal with $\omega_{RF}$ is intensity-modulated onto the idler photons, whose transfer function is given by $M(t_2) = 1 + \cos(\omega_{RF}t_2)$, the temporal waveforms of the idler photons conditioned on the temporal correlation can be deduced as

$$\rho_c(t_2) = \int dt_1 \Psi'(t_1,t_2)M(t_2) \propto e^{-\frac{\widetilde{\omega}_F^2}{2\sigma_F^2}}\big(1+\cos(\omega_{RF}t_2)\big). \quad (6)$$

Without loss of generality, the global term of $e^{-\widetilde{\omega}_F^2/2\sigma_F^2}$ is neglected in the following deduction. Thus, the temporal waveform of the conditioned idler photons can be rewritten as $\rho_c(t_2) \propto 1+\cos[\omega_{RF}t_2]$. Likewise, the temporal waveforms of the conditioned signal photons can be deduced as

$$\rho_c(t_1) = \int dt_2 \Psi'(t_1,t_2)M(t_2)$$
$$\propto 1+\frac{1}{2}e^{\frac{\widetilde{\omega}_F^2}{2\sigma_F^2}}\left(e^{-\frac{(\widetilde{\omega}_F-\omega_{RF})^2}{2\sigma_F^2}}e^{-i\omega_{RF}t_1}+e^{-\frac{(\widetilde{\omega}_F+\omega_{RF})^2}{2\sigma_F^2}}e^{i\omega_{RF}t_1}\right) \quad (7)$$

For $\widetilde{\omega}_F \gg \omega_{RF}$, Eq. (7) can be approximated as $\rho_c(t_1) \propto 1+\cos[\omega_{RF}t_1]$, which takes the same form with Eq. (6). Thus, the RF modulation on the idler photons is nonlocally transferred onto the signal photons based on the photon conditioning. The temporal modulation and mapping are also plotted in Fig. 1b.

For the implementation of nonlocal RF phase shifting, a dispersive element with a dispersion parameter of $D$ should be input into the setup. Consider it is on the idler photon path, the corresponding transfer function can be given by $H_D(t_2) = \exp\left[-i\left(\frac{t_2^2}{2D}+t_2\widetilde{\omega}_F\right)\right]$. Without involving the RF modulation, the resultant two-photon temporal wave function can be given by

$$\Psi''(t_1,t_2) \propto \int d\tau\, \Psi'(t_1,\tau)H_D(\tau-t_2) \propto \exp\left[\frac{i(t_1-t_2)^2\frac{\sigma_F^2}{2}-iD\widetilde{\omega}_F\sigma_F^2(t_1+t_2)+\frac{D}{2}\widetilde{\omega}_F^2+2\widetilde{\omega}_F t_2}{\mathrm{i}+2D\sigma_F^2}\right], \quad (8)$$

whose square module is then deduced as

$$|\Psi''(t_1,t_2)|^2 \propto e^{-\frac{(t_1-t_2-D\widetilde{\omega}_F)^2}{1/\sigma_F^2+D^2\sigma_F^2}}. \quad (9)$$

It can be seen that, the dispersion $D$ will introduce a center shift dependent on the slicing frequency ($\tau = -\widetilde{\omega}_F D$) of the temporal correlation distribution besides the width broadening, whose profile is also shown in Fig. 1c by the green shape. With the RF modulation on the idler photons before the dispersion, the temporal waveform of the idler photons conditioned on the temporal correlation can be deduced as

$$\begin{aligned}\rho_c'(t_2) &= \int dt_1 \int d\tau\, \Psi'(t_1,\tau) M(\tau) H_D(\tau - t_2) \\ &\propto \int dt_1\, e^{-i\widetilde{\omega}_F(t_1-t_2)} e^{-\frac{i\sigma^2(t_1-t_2)^2}{2+2\sigma^2 D}} \left\{1 + \frac{1}{2}\begin{pmatrix} e^{i\omega_{RF}t_1} e^{\frac{\omega_{RF}(t_1-t_2-\omega_{RF}D/2)}{i+D\sigma^2}} \\ +e^{-i\omega_{RF}t_1} e^{-\frac{\omega_{RF}(t_1-t_2+\omega_{RF}D/2)}{i+D\sigma^2}} \end{pmatrix}\right\} \\ &\propto 1 + e^{iD\omega_{RF}^2/2} \cos[\omega_{RF}(t_2 - D\widetilde{\omega}_F)] \end{aligned} \quad (10)$$

It can be seen that, the conditioned idler photon waveform experiences a phase shift of $\varphi = \omega_{RF}\tau = -\omega_{RF}\widetilde{\omega}_F D$. The term $e^{iD\omega_{RF}^2/2}$ represents the microwave frequency-dependent fading, which is analogous to the classical MWP technoloty. However, the fading associated with the carrier bandwidth in the classical MWP has no influence at all. The corresponding temporal waveform of the signal photons conditioned on the temporal correlation can also be deduced, which is given by

$$\begin{aligned}\rho_c'(t_1) &= \int dt_2 \int d\tau\, \Psi'(t_1,\tau) M(\tau) H_D(\tau - t_2) \\ &\propto 1 + \frac{1}{2} e^{\frac{\widetilde{\omega}_F^2}{2\sigma_F^2}} \left( e^{-\frac{(\widetilde{\omega}_F - \omega_{RF})^2}{2\sigma_F^2}} e^{-i\omega_{RF}t_1} + e^{-\frac{(\widetilde{\omega}_F + \omega_{RF})^2}{2\sigma_F^2}} e^{i\omega_{RF}t_1} \right) \end{aligned} \quad (11)$$

It can be seen that, Eq. (11) takes the same form with Eq. (7). Therefore, the nonlocally mapped RF modulation on the signal photons will not be affected by the dispersion in either phase shift or fading[24]. The expected phase shifting performances for both the signal and idler photon waveforms are plotted in Fig. 1c. Note should be taken that, when a dispersive element of $D'$ is inserted into the signal path, the results should be vice versa. That is, the conditioned signal photon waveform should experience a phase shift of $\varphi = \omega_{RF}\widetilde{\omega}_F D'$ and a fading determined by the term $e^{iD'\omega_{RF}^2/2}$.

When a multi-channel optical slicing with identical spectral gap of $\Delta\widetilde{\omega}$ and spectral bandwidth of $\sigma_F$ is applied onto the signal photons, the transfer function can be given by $H'(\widetilde{\omega}_s) \sim \sum_k \exp\left[-\left(\widetilde{\omega}_s - \widetilde{\omega}_{F,0} - k\Delta\widetilde{\omega}\right)^2/(2\sigma_F^2)\right]$. Due to the frequency anticorrelation between the signal and idler photons, the spectral slicings are nonlocally mapped onto the idler photons as shown in Fig. 1d. Similar to the deductions from Eq. (3)-(11), the temporal waveform expression of the idler photons conditioned on the temporal correlation can be given by

$$\rho_c''(t_2) \propto 1 + e^{iD\omega_{RF}^2/2} \sum_k \cos\left[\omega_{RF}\left(t_1 - D\left(\widetilde{\omega}_{F,0} + k\Delta\widetilde{\omega}\right)\right)\right] \quad (12)$$

As each spectrally sliced idler photons would experience a different phase shifting linearly dependent on the dispersion as shown in Fig. 1e, the acquired photon waveforms is determined by the function $\sum_k \cos\left[\omega_{RF}\left(t_2 - D\left(\widetilde{\omega}_{F,0} + k\Delta\widetilde{\omega}\right)\right)\right]$ which then results in the multi-tap transversal frequency filter with the free-spectral range ditermined by $\Delta\widetilde{\omega}D$. In comparison with the classical MWP, no fading associated with the carrier bandwidth contributes. Likewise, with another dispersive element with a dispersion of $D'$ simultaneously inserted into the signal photon path as shown in Fig. 1f, the acquired photon waveform from the signal photon flow should follow

$$\begin{aligned}&\rho_c''(t_1) \propto 1 + \frac{1}{2} e^{iD'\omega_{RF}^2/2} \sum_k e^{\frac{(\widetilde{\omega}_{F,0}+k\Delta\widetilde{\omega})^2}{2\sigma_F^2}} \begin{pmatrix} e^{-\frac{(\widetilde{\omega}_{F,0}+k\Delta\widetilde{\omega}-\omega_{RF})^2}{2\sigma_F^2}} e^{-i\omega_{RF}\left(t_1 + D'\widetilde{\omega}_F\right)} \\ +e^{-\frac{(\widetilde{\omega}_{F,0}+k\Delta\widetilde{\omega}+\omega_{RF})^2}{2\sigma_F^2}} e^{i\omega_{RF}(t_1+D'\widetilde{\omega}_F)} \end{pmatrix} \\ &\xrightarrow{\widetilde{\omega}_{F,0}+k\Delta\widetilde{\omega} \gg \omega_{RF}} \rho_c''(t_1) \propto 1 + e^{iD'\omega_{RF}^2/2} \sum_k \cos\left[\omega_{RF}\left(t_1 + D'\left(\widetilde{\omega}_{F,0} + k\Delta\widetilde{\omega}\right)\right)\right] \end{aligned} \quad (13)$$

Thus, a second multi-tap frequency filter can be realized from the signal photon output with a different free-spectral range ditermined by $\Delta\widetilde{\omega}D'$. Note that, in practical experiments, the spectral slicing is defined in the wavelength domain. By transforming $\widetilde{\omega}_F$ into the wavelength deviation ($\delta\lambda_F$) from the degenerate wavelength ($\lambda_0$) and the dispersion $D$ into the group delay dispersion (GDD), the time delay can be rewritten as $\tau \approx \delta\lambda_F \cdot GDD$. Accordingly, the FSR should be given by $FSR \approx 1/(\Delta\lambda \cdot GDD)$, where $\Delta\lambda$ corresponds to the wavelength gap between the adjacent spectral slices.

## Experimental Setup

The experimental setup of the QMWP signal processing system is shown in Fig. 2. The frequency anti-correlated and temporal correlated photon pairs are generated from a piece of 10 mm long, type-II PPLN waveguide (HC Photonics) pumped by a CW distributed Bragg reflector laser (DBR laser Photodigm) at 780 nm, whose spectral bandwidth in FWHM was measured as 0.0025 nm[28]. After filtering out the residual pump beam, the orthogonally polarized signal and idler photons, marked with $s$ and $i$, are spatially separated by a fiber polarization beam splitter (FPBS). The FWHM single-photon spectral widths for both the signal and idler photons were measured as about 2.4 nm[28]. To implement the optical slicing, a programmable waveshaper (WaveShaper 16000A, Finisar) is placed in the signal path. For the QMWP phase shifting (Fig. 2b) demonstration, the shaper is set to have a single passband with a tunable central wavelength. For the QMWP 3-tap transversal filtering (Fig. 2c), the shaper is set to have 3 passbands with uniform bandgap and tunable attenuation. In the idler path, the photon flux is intensity modulated by a Mach-Zehnder modulator (MZM, PowerBit™ F10-0, Oclaro), through which the high-speed RF signal from a signal generator (E8257D, KeySight) is loaded onto the idler photons. Then the modulated idler photons are fed into a fiber-Bragg-grating-based dispersion compensation module (DCM, Proximion AB) for realizing the dispersion-induced phase shifting. In the experiment, different DCMs with the group delay dispersion (GDD) values ranging from 165 ps/nm to 495 ps/nm are applied. Afterward, the processed signal and idler photons are respectively detected by the low-jitter superconductive nanowire single-photon detectors (SNSPD1 & SNSPD2, Photec) with the time jitter of about 50 ps[29]. The two SNSPD outputs are then delivered to different input ports (CH1 and CH2) of the TCSPC module (PicoQuant Hydraharp 400), which is operated in the Time Tagged Time-Resolved (TTTR) T3 mode with its time-bin resolution being set as 8 ps. In the T3 mode, the TCSPC records the arrival times of the photon events for each input port and sync events for the "Sync" port. The 10 MHz time base from the signal generator E8257D is used for establishing phase stabilization between the RF signal and the sync signal. Based on the coincidence measurements, the entangled photon pairs are post-selected and utilized to rebuild the waveforms processed by the QMWP method[24].

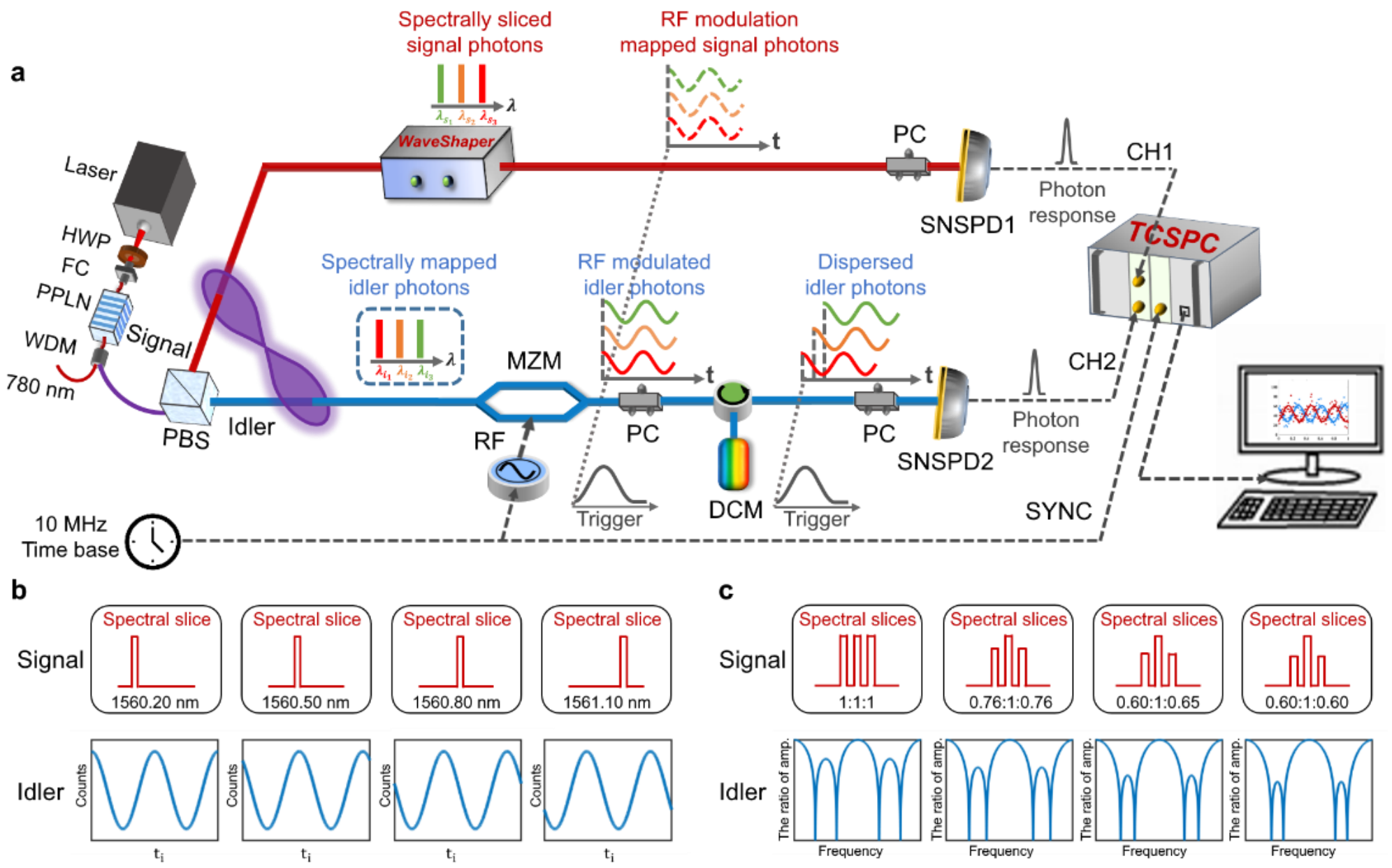

**Fig. 2 Experimental setup of the QMWP RF frequency phase shifter and filter. a** The photon pairs are from the process of SPDC. The signal photon is optically sliced by a programmable wave shaper, while the idler photon is modulated and dispersed. Afterwards, the signal and idler photons are detected by a multichannel SNSPD and a TCSPC for recording their time of arrival. By applying the cross-correlation searching algorithm, the paired signal and idler photons are selected out and utilized to reconstruct the photon waveforms. **b** The graphics of the idler photon waveforms when the wavelength channel of the single-passband optical slice is increased, which shows the QMWP functionality of RF phase shifting. **c** The graphics of the idler photon waveforms when the three passbands of the optical slicer take different ratios, which illustrates the QMWP functionality of 3-tap filter with different sidelobe suppression ratios.

## Experimental result

### 1.QMWP-based RF signal phase shifting performance

The nonlocal RF signal phase shifting on the idler photons is first investigated based on the setup. The waveshaper is set in the single-passband mode, with its FWHM bandwidth being 0.30 nm and the center wavelength being tuned from 1560.20 nm to 1561.10 nm with a step of 0.30 nm. When the RF signal is 3 GHz and the DCM in the idler path is chosen to have a GDD of 165 ps/nm, the reconstructed waveforms for the signal (red dots) and idler (blue squares) photons at the four successive wavelengths are plotted in the four subfigures of Fig. 3a. It can be seen that the waveforms of the non-dispersed signal photons do not vary with the wavelength adjustment, while the waveforms of the dispersed idler photons experience a left-hand shift with the increasing of the wavelength. With the DCM moved to the signal path, similar measurements are also made and plotted in Fig. 3b. By contrast, the idler photon waveforms do not vary with the wavelength while the waveforms from the signal photons experience a right-hand shift. By extracting the relative phase difference between the idler and signal photon waveforms over wavelength, the system's phase-shifting performance can be evaluated and depicted in Fig. 3c. From the linear fittings to the phase differences, which give the slopes of $3.27\pm0.07$ rad/nm and $3.31\pm0.03$ rad/nm respectively, the simulated GDD values of $173.48\pm3.71$ ps/nm and $175.60\pm1.59$ ps/nm show good consistence between each other as well as with the nominal value. For better verification, further tests are applied by applying a lager GDD of 330 ps/nm (Fig. 3d) and higher RF-frequency signal of 5 GHz (Fig. 3e). Nice agreements with the theoretical expectation are also achieved for the two cases, indicating that the QMWP system can well perform the high-speed RF phase shifting.

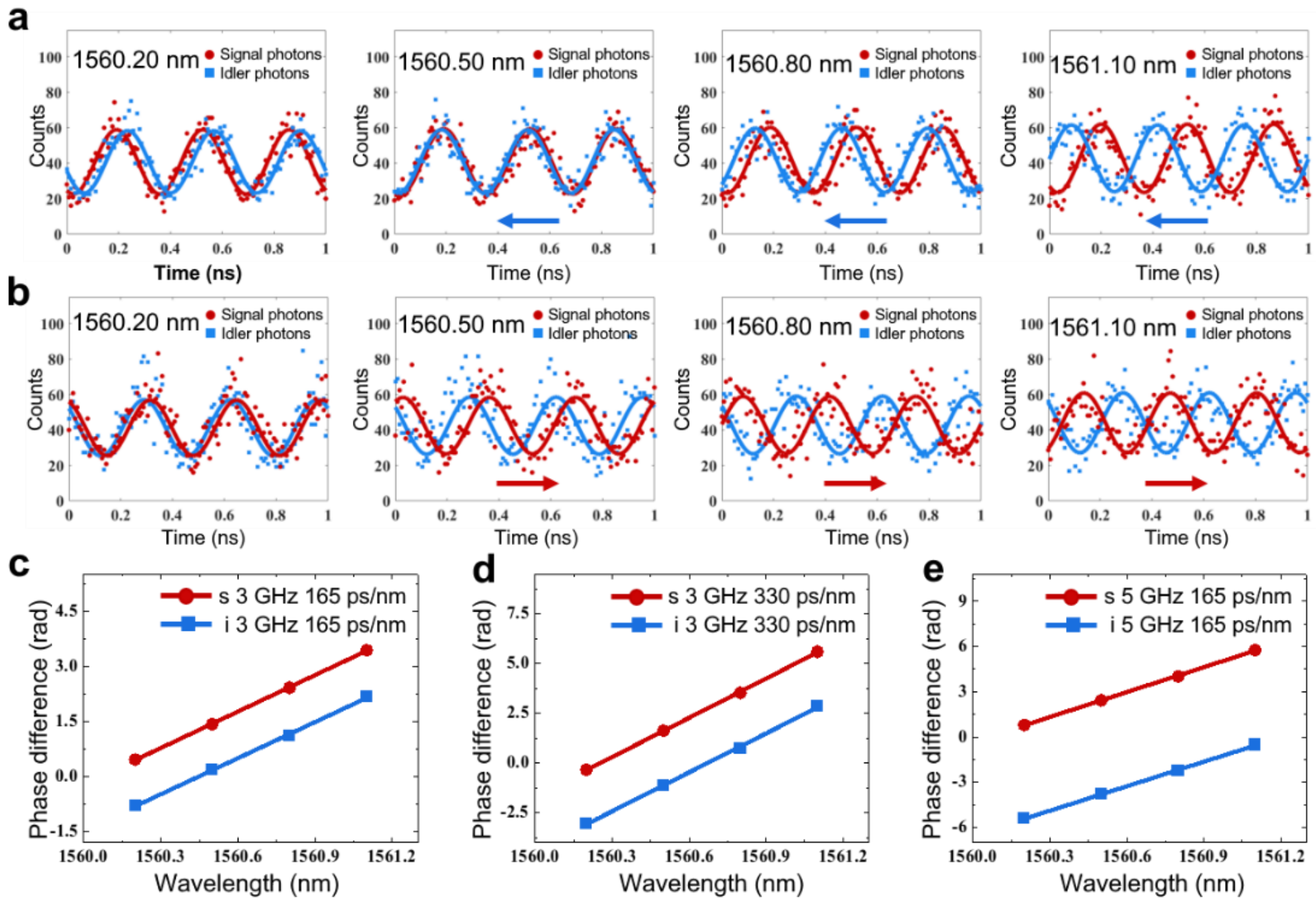

**Fig. 3 Results of the QMWP-based RF signal phase shifting. a** The reconstructed temporal waveforms of the signal and idler photons (plotted in red dots and blue dots respectively) when the DCM with a GDD of 165 ps/nm is placed in the idler path and a 3 GHz RF signal is applied. With the adjustment of the central wavelength of the tunable filter, the waveform of the signal photon remains unchanged, while the waveform of the idler photon is left-hand shifted. **b** The reconstructed temporal waveforms of the signal and idler photons when the DCM moves to the signal path while all the other settings are maintained. In this case, the waveform of the signal photon is right-hand shifted, while the waveform of the idler photon remains unchanged. **c** The extracted relative phase differences over wavelength between the signal and idler photons' waveforms in **a** (blue squares) and **b** (red dots). The slopes of the linear fits are 3.27 ±0.07 rad/nm and 3.31 ±0.03 rad/nm, respectively. According to the direct dependence on the product of $\omega_{RF}$ and GDD, the measured GDDs are 173.48 ±3.71 ps/nm, 175.60 ±1.59 ps/nm, which are in good consistence and close to the nominal value of 165 ps/nm GDD. **d** and **e** demonstrate the phase shifting performance when the RF-frequency is set to 5 GHz and the GDD of the DCM is changed to 330 ps/nm, respectively. The measured GDDs in **d** and **e** are respectively 346.43 ±2.23 ps/nm and 349.61 ±2.86 ps/nm, 172.52 ±4.77 ps/nm, 175.39 ±4.24 ps/nm, indicating the nice agreements with nominal GDD values of the DCMs.

## 2.QMWP-based RF transversal filtering performance

### 2.1 Comparison between the QMWP method and the SP-MWP method

As has been pointed out in the Principle section, the QMWP-based RF transversal filtering performance can be realized when the optical slicer has multiple wavelength passbands with a uniform bandgap. By setting the passbands to be three with their center wavelengths being 1560.95 nm, 1561.75 nm, and 1562.55 nm respectively and each having a FWHM bandwidth of 0.3 nm, the 3-tap transversal filtering performance in the QMWP configuration is tested. The power ratio of the three passbands can be flexibly adjusted. By adjusting the power ratio as 1:1:1, the three passbands with the same output power are utilized as the subsequent optical carrier. With the help of the MZM, the RF signal with a modulation power of 10 dBm and an adjustable frequency from 200 MHz to 8 GHz can be uploaded on the three optical carriers. Based on the post-selected idler photon counts, the reconstructed waveforms over time with respect to the two cases with and without the DCM in the setup are measured. For the two DCMs cases with GDD of 330 ps/nm and 495 ps/nm, the amplitude ratios of the waveform with dispersion to that without dispersion as a function of the RF from 200 MHz to 8 GHz are respectively investigated (see Supplementary information) and plotted in Fig. 4a and 4b by orange diamonds. Both show a nice-shaped notch filtering pattern with their free spectral range (FSR) being 3.50 GHz and 2.35 GHz respectively. Taking the dispersion-induced frequency fading effect without carrier bandwidth contribution (solid yellow curves) into consideration, the simulations of the filtering performances for both cases are given by the solid blue curve, which shows a nice agreement with the results.

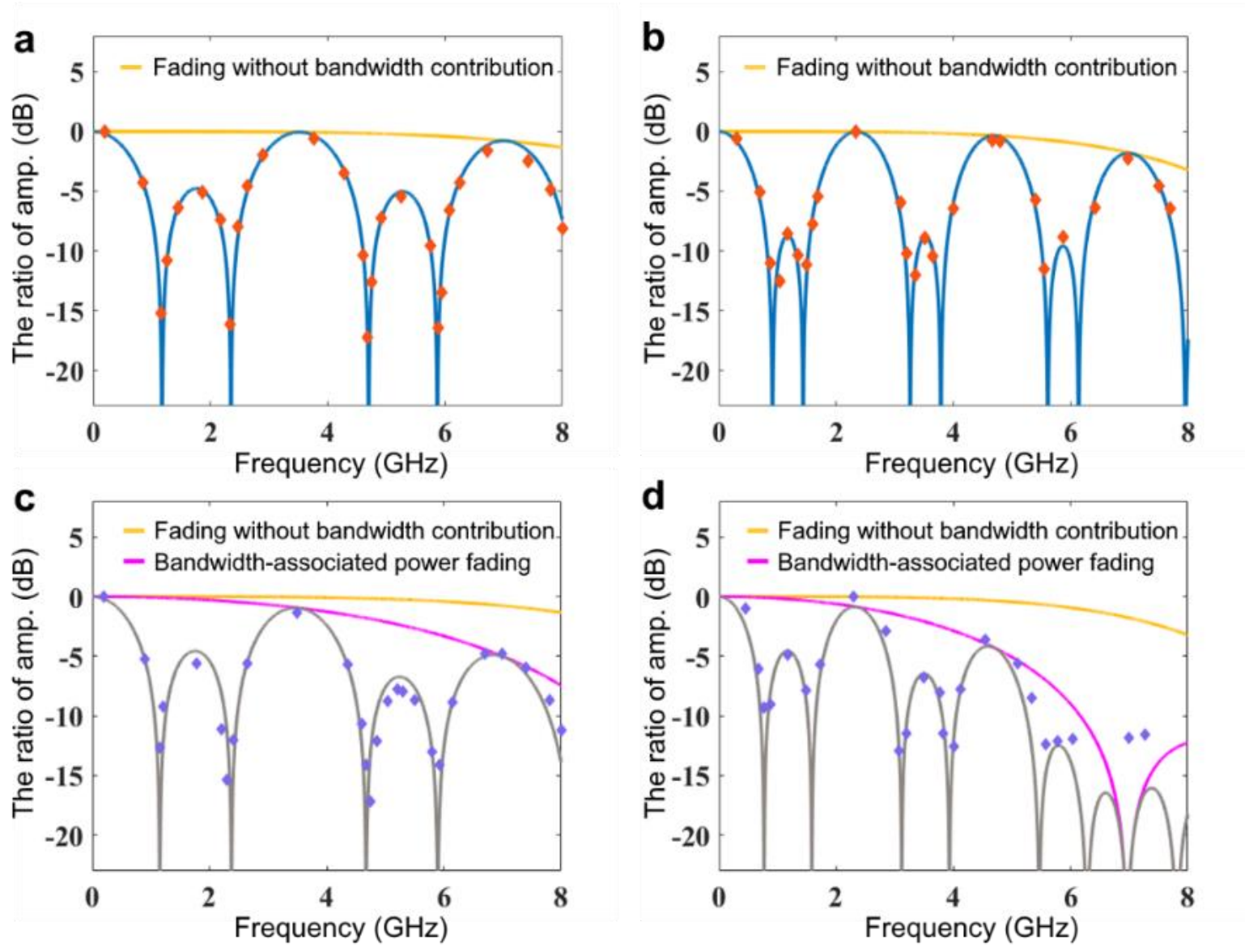

**Fig. 4 The comparison between QMWP and SP-MWP RF frequency filtering.** The experimentally acquired 3-tap filtering responses in the configurations of QMWP (**a** 330 ps/nm and **b** 495 ps/nm) and SP-MWP (**c** 330 ps/nm and **d** 495 ps/nm). The three-tap RF filtering performance is tested by programming the waveshaper to have 3 passbands, whose center wavelengths are 1560.95 nm, 1561.75 nm, and 1562.55 nm, respectively and each has an FWHM bandwidth of 0.3 nm. The solid yellow curve depicts the dispersion-induced frequency fading effect without carrier bandwidth contribution, and the solid pink curve in **c** and **d** involve the dispersion-induced frequency fading effect associated with the carrier bandwidth (0.3 nm in our case).

To reveal the superiority of the QMWP method for realizing the RF filtering, the 3-tap RF filtering based on the SP-MWP method is also implemented by arranging all the above-mentioned elements of wave shaping, MZM, and DCM in the idler photon path. At different RF in the range of 200 MHz to 8 GHz, the TCSPC-recovered photon waveforms over time from the detected *i* path photon events are measured with respect to the two cases with and without the DCM in the setup. Regarding to the above two DCM cases, the amplitude ratios of the waveforms with dispersion to that without dispersion as a function of the RF from 200 MHz to 8 GHz are also plotted in Fig. 4c and 4d by purple diamonds. As can be seen, the ratio decrements with the increasing RF are much faster than the dispersion-induced frequency fading without carrier bandwidth contribution (solid yellow curves) and should be due to the carrier bandwidth-associated power fading. By taking the 0.3 nm bandwidth of the carriers into account (solid pink curves), a nice fitting can be found between the simulations (solid gray curves) and results (purple diamonds) in Fig. 4c and 4d. The results demonstrate that, the frequency transversal filtering based on the SP-MWP method, whose performance is equivalent to the classic MWP with the broadband optical carrier, is easily affected by the dispersion-induced fading associated with carrier bandwidth. Although larger dispersion would lead to narrower filtering bandwidth, it would also result in worse fading. On the contrary, the QMWP-based RF transversal filtering performance (as shown by Fig. 4a and 4b) demonstrates a perfect immunity to the fading associated with the carrier bandwidth.

### 2.2 The reconfigurability of the 3-tap RF filter based on the QMWP method

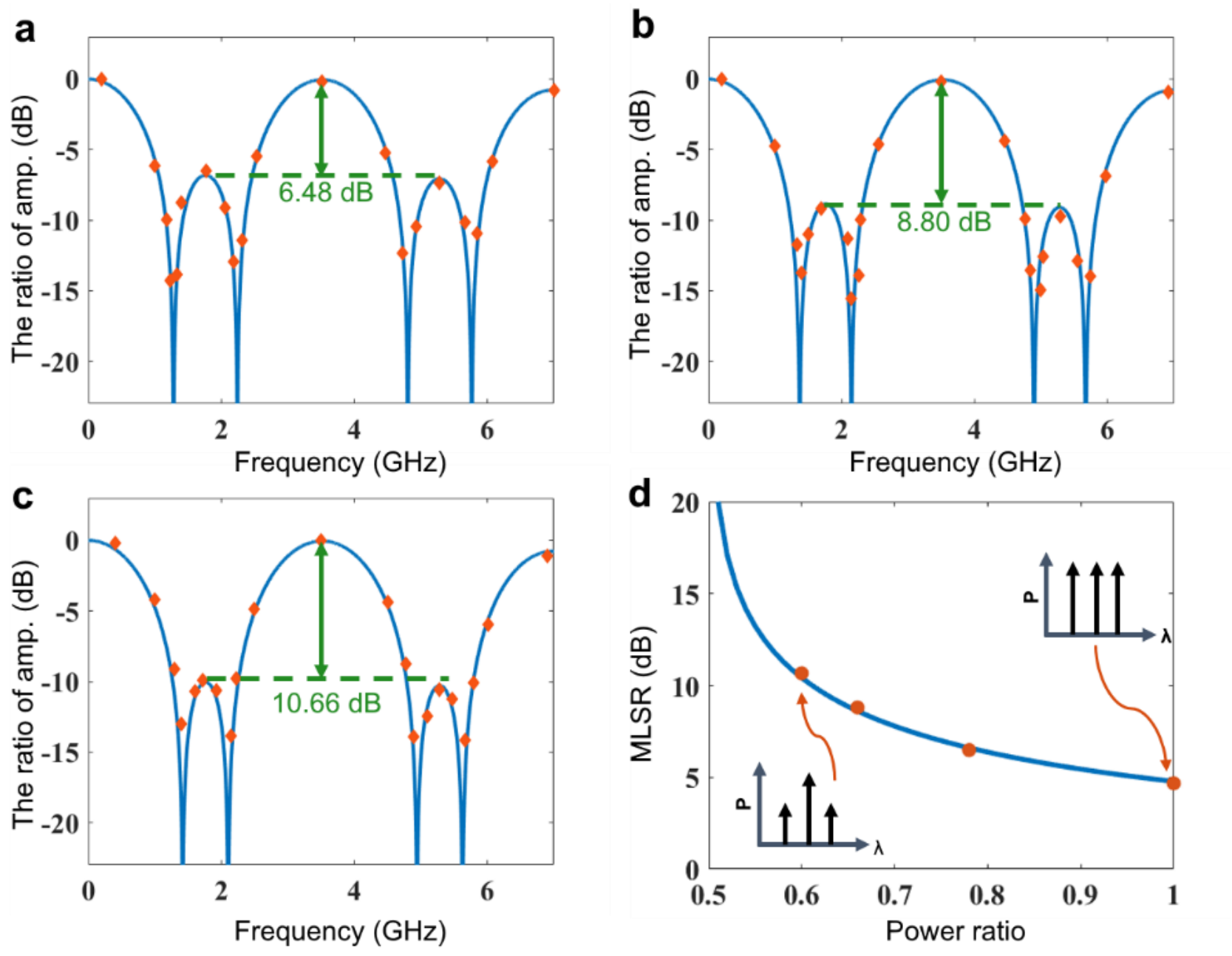

**Fig. 5 The reconfigurability of the 3-tap QMWP RF filter.** The 3 passbands of the waveshaper are maintained at the center wavelengths of 1560.95 nm, 1561.75 nm, and 1562.55 nm respectively. Reconfigurable sidelobe suppression of the 3-tap RF-filter is investigated via the filtering responses with the GDD being 330 ps/nm while the 3 passbands photon flux ratio is taken as **a** 0.76:1:0.76 **b** 0.65:1:0.65, and **c** 0.60:1:0.60 respectively, which correspond to different sidelobe suppression ratio of a 6.48 dB **b** 8.80 dB, and **c** 10.66 dB. **d**. The comparison between the measured MLSR of QMWP nonlocal filtering and the simulation of the classic MWP filtering.

In addition to the dispersion immunity, the QMWP transversal filter inherits the excellent tunability both in the FSR (see Supplementary Note 3) and in the sidelobe suppression capability. The sidelobe suppression capability of the three-tap RF filter is evaluated by measuring the main to sidelobe ratio (MSLR). The 3 passbands of the waveshaper are maintained at the center wavelengths of 1560.95 nm, 1561.75 nm, and 1562.55 nm respectively. Reconfigurable sidelobe suppression of the 3-tap RF-filter is investigated and depicted in Fig. 5 with the GDD being 330 ps/nm while the 3 passbands photon flux ratio is taken as **a** 0.76:1:0.76, **b** 0.65:1:0.65, and **c** 0.60:1:0.60 respectively, which give different MSLR of **a** 6.48 dB, **b** 8.80 dB, and **c** 10.66 dB. Together with the result shown in Fig. 4a, which gives a MSLR of 4.67 dB, it is clearly manifested that the MSLR can be improved by decreasing the photon flux ratio of the sideband channels to the center channel. The dependence of the MSLR on the flux ratio is then plotted in Fig. 5d by orange dots, and the simulation based on the classical MWP theory is also given by the solid blue curve. The excellent agreement with the experimental results demonstrated that the filtering in the QMWP follows the same filtering performance with the classical MWP method.

### 2.3 Parallel RF transversal filtering with two-dimensional outputs

As already noted before, though the above demonstrations are implemented with the DCM in the idler path, the equivalent QMWP signal processing performance should be expected when the DCM is moved to the signal path (see Supplementary Note 4). Therefore, when each of the signal and idler paths is applied with a DCM, the QMWP 3-tap filter becomes a parallel RF processor which can provide two-dimensional outputs with different FSR responses. Figure 6 demonstrates the two-dimensional 3-tap QMWP filter by placing a DCM with the GDD of -330 ps/nm in the signal path while a DCM of 165 ps/nm in the idler path. It can be seen that, the filtering FSR responses in the signal path (Fig. 6a) and the idler path (Fig. 6b) are different. The result further verifies that, the dispersion, unlike the spectral selection and temporal modulation, cannot be directly mapped from one photon to its twin entangled photon.

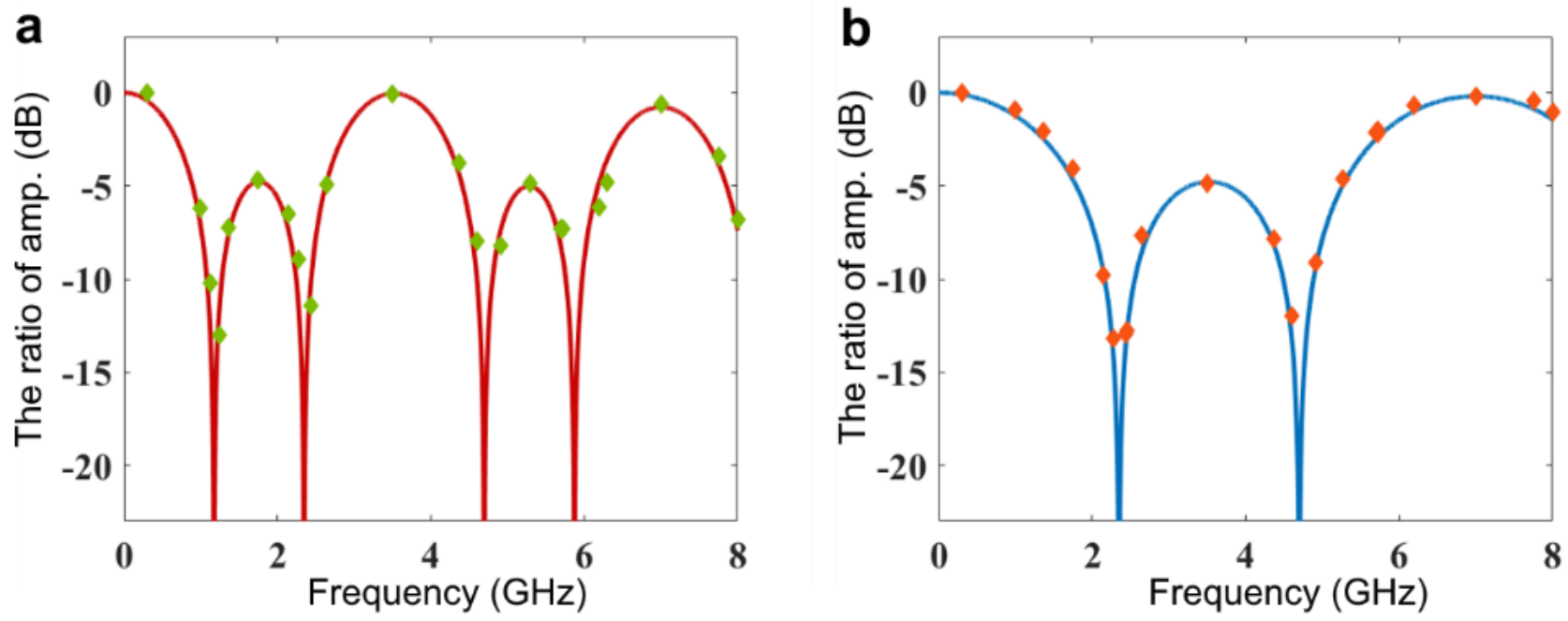

**Fig. 6 The two-dimenional parallel QMWP RF-filtering demonstration.** In the experimental setup, the DCM of -330 ps/nm is placed in the signal path while the DCM of 165 ps/nm is placed in the idler path: **a** is the RF-filtering result extracted from the signal photon waveform, and **b** is the RF-filtering result extracted from the idler photon waveform.

## Discussion

In summary, we have demonstrated a QMWP signal processing system with the energy-time entangled photon pairs being the optical carrier. Benefiting from the frequency anticorrelation and temporal correlation between the photon pairs, the RF modulation and spectral slicing on one photon can be nonlocally mapped onto its 'twin' one. Based on the such quantum-facilitated features, nonlocal RF phase shifting and multi-tap frequency filter can be implemented. Such capability is crucial to realize encrypted microwave processing in modern communication and radar applications. More importantly, the QMWP RF filtering presents the perfect immunity to the dispersion-induced frequency fading effect associated with the broadband carrier, which is still a big challenge in classical microwave photonics. Furthermore, as both the signal and idler photon events can be utilized to extract the RF signal, a native two-dimensional parallel microwave signal processor can be provided. For example, the QMWP transversal filter has shown two parallel outputs with different FSR responses. These demonstrated features fully prove the superiority of QMWP over classical MWP and open the door to numerous possible applications of microwave photonics and related fields.

## Methods

**The calculation of RF filtering frequency response.** Since the TCSPC only records the temporal information of the detected photon, the temporal waveform of the modulated signal reconstruction needs further processing. Therefore, traditional spectral measuring equipment like spectral analyzers and vector network analyzers cannot be used. By comparing the amplitude of the reconstructed temporal waveform with and without the dispersion at different input signal's frequencies, the frequency response of the QMWP RF filter can be calculated. This is the reason for using the amplitude ratio to demonstrate the RF filtering performance. Moreover, the amplitude of the reconstructed temporal waveform without the dispersion can be calculated from the waveform of its non-dispersed twin photon because of the temporal correlation of the photon pairs. For example, If the dispersion module is only placed in the idler path, the waveform of the non-dispersed signal photon will be the same as the waveform reconstructed from the idler photon without dispersion. If only one photon in the photon pair is dispersed, the frequency response is obtained by calculating the amplitude ratio between the waveform of the dispersed photon and the waveform of the non-dispersed twin photon. If both photons in the photon pair are dispersed, an extra path separated before DCMs is needed to calculate the amplitude of the reconstructed waveform without the dispersion.

### Funding.

The National Natural Science Foundation of China (Grant Nos. 12033007, 61875205, 91836301, 12103058, and 61801458), the Key Project of Frontier Science Research of Chinese Academy of Sciences (Grant No. QYZDB-SSW-SLH007), the Strategic Priority Research Program of CAS (Grant No. XDC07020200), the Youth Innovation Promotion Association, CAS (Grant No. 2021408, 2022413), the Western Young Scholar Project of CAS (Grant Nos. XAB2019B17 and XAB2019B15), the Chinese Academy of Sciences Key Project (Grant No. ZDRW-KT-2019-1-0103).

### Acknowledgments

We thank Honglei Quan and Wenxiang Xue for their useful technical discussions.

### Author contributions

R.D., Y.J., M.L., and S.Z proposed this project. R.D. built the theory model. Y.J. designed and performed the experiment. Y.J, Y.Y. and H.H. carried out numerical simulations and analyzed the data. Y.Y, H.H, X.X, R.Q, T.L, N.Z., and M.L contributed to the discussion of experimental results. Y.J., R.D., and S.Z. wrote the paper. M.L., R.D., and S.Z supervised the research work. All the authors commented and revised the paper.

### Disclosures.

The authors declare no conflicts of interest.

### Data availability.

The data that support the findings of this study are available from the corresponding author upon reasonable request.